# Answers to a few questions regarding the BMV experiment

*Chiara Marletto & Vlatko Vedral*
*Clarendon Laboratory, University of Oxford*
*July 2019*

**ABSTRACT**: We provide brief answers to a number of recurring questions about the BMV effect and the related experimental proposal (Bose *et al.*, 2017; Marletto & Vedral, 2017). Some of these questions include alleged counter-examples to our result, stating that if gravity (or anything mediating the interaction) can create entanglement while complying with locality, then it must then be non-classical. Here we explain why these objections are not counter-examples and why our result still stands; we also explore a few other important subtleties often disregarded on a first read of our paper (Marletto & Vedral, 2017).

**Summary of the result.** The Bose-Marletto-Vedral effect (BMV effect for short hereafter) can be summarised as follows. First, one shows, with a general argument, that if two quantum systems **Q1** and **Q2**, located at spatially separate locations 1 and 2, become entangled through the local mediation of a third system M, which interacts locally at 1 with **Q1**, and at 2 with **Q2**, then M has to be 'non-classical', in the following sense: **M** has to have at least two variables that cannot be simultaneously measured by the same machine to an arbitrarily high accuracy. More formally, these are two variables representable each by two operators that do not commute with one another. Hence M may or may not be a fully quantum system (Hall & Reginatto, 2017), but if it can entangle **Q1** and **Q2** by locally mediating their interaction, one can conclude that it cannot be described by a classical theory only. Based on this argument, one can propose a realistic experiment where **Q1** and **Q2** are two masses, and **M** is the gravitational field. Using a quantum model based on linear quantum gravity, one can then show that there is an experimentally accessible regime, where maximal path entanglement can be formed between **Q1** and **Q2** using gravity as the only mediator. The proposed experiment, currently under consideration for further feasibility studies, is intended to detect this entanglement. A generally covariant description for this model can be found in (Christodoulou&Rovelli, 2019).

**Questions and Answers.**

We present the usual questions we receive on our work below, immediately followed by the answers. Our exposition will be deliberately simple and without any detailed mathematics in order to reach as wide an audience as possible (we will be posting a more detailed follow-up in due course).

1. **Question:** I have two atoms in my optical trap. I shine classical laser light on both and they become entangled. Therefore, classical light can entangle two atoms, and so classical gravity should be able to do the same with the masses in your proposed experiment! How can this be compatible with your result, which says that anything capable of entangling two quantum systems must itself be quantum?

**Answer:** Our result is more precise than that. It states that anything capable of entangling two quantum systems AND SATISFYING LOCALITY (plus a few other assumptions) must itself be quantum. The locality assumption means that the two objects to be entangled should not be interacting directly, but only locally, at their respective locations, with the mediator. This key assumption is violated in the above example. Classical light is a stream of (lots and lots of) photons whose wavelength is bigger than the separation between the atoms. When a photon is absorbed, the information about which atoms has absorbed it is therefore not available. They thus create an entangled state where one atom is excited and the other not and vice versa. This fact does not contradict our theorem because the mediator (the photon) is highly delocalized and the interaction between each atom and the photon is not localized at the points of the atoms, contrary to the locality assumption. In the proposed experiment, it is crucial to exclude any other means for the two masses to get entangled other than by locally interacting with the gravitational field.

2. **Question:** I have a BBO type II crystal on my optical bench. I shine some classical light onto it. Every once in a while, I get two entangled photons out. Therefore, classical light and a classical piece of glass can create entanglement.
   **Answer:** No, they cannot. There are two points relevant here. One is that the BBO crystal is made up of lots and lots of molecules. Their internal structure is such that when they absorb a photon, there is a probability that they emit two photons (two transitions in a rapid succession) which are entangled. These transitions are fully quantum, in that the electron pertaining to the molecule goes through all the superposed states of the energies while emitting. If it could not do so, there would be no emission. Therefore, the generation is a fully quantum process. The second point is that the interaction leading to the entanglement happens again in a way that is non-local, in that the whole process occurs in individual molecules (all in a small spatial volume). If so, then as explained earlier, our argument does not apply, because of the violation of the locality condition.

3. **Question:** I have a beam-splitter in my lab. I send a single photon in, and out comes a superposition of the photon in two output ports. This can be thought of as an entangled state between the two output modes. Therefore, a classical beam-splitter can create entanglement, again, contrary to your result.
   **Answer:** Yes, it can, but this does not invalidate our result –again because in this case the assumption about the locality of interactions fails. The interaction between the modes which get entangled is directly happening at one point, where the photon is, as the photon meets the beam-splitter, therefore it is mediated by atoms/molecules in the beam-splitter at that point (much as in the molecule of the BBO crystal above). This is not the same as having two distant modes that are not interacting with one another directly, whose interaction is mediated by an intermediate field. Therefore, no contradiction arises between our conclusion and beam-splitters creating single photon entanglement.

4. **Question:** In the linear model of quantum gravity, the Newtonian interaction in the Coulomb gauge is an instantaneous static term. Therefore, the entanglement between two masses is generated by an entirely classical gravitational interaction (see C. Anastopoulos and B.-L. Hu).

   **Answer:** This is untrue. In the Coulomb gauge, when two masses are static, Newtonian gravity is indeed just an instantaneous classical term in the Hamiltonian. However this is insufficient to describe our effect. In the BMV effect, there is a dynamical component, due to the fact that one has to form a spatial superposition of each mass and then undo it, for interference to take place. In this dynamical (albeit slowly-varying) scenario, the transverse quantum degrees of freedom become crucial as the mediators of the interaction. The interaction between the two is mediated by the transverse degrees of freedom and the final path entanglement is therefore only possible because the mediating field is quantum mechanical and provides those (non-commuting) degrees of freedom. The interaction is local, in the sense that each mass couples locally to the (gravitational) field which then propagates between the two masses. This is the same scenario as when we model Coulomb interaction between two moving charges in quantum electrodynamics.

5. **Question:** This result, that if a mediator can entangle two other quantum systems, then it has to be quantum, is obvious! It is equivalent to the quantum information theorems about impossibility of creating entanglement via local operations and classical communication (LOCC).

   **Answer:** This is also untrue. The subtle, but essential, point here is that the mediator **M** should not be assumed to obey quantum theory (otherwise, the argument is close to being circular; it only discriminates separable form entangled, but all quantum, states). Therefore, the standard argument about LOCC don't apply at all. The argument supporting the experiment has to be cast in a scenario that does not assume quantum theory to hold for the three systems (the two quantum masses and the mediator), only assuming more general principles (e.g. locality, rules constraining observables – see Marletto&Vedral, 2017 and Marletto&Vedral, 2016).

**References.**

1. C. Anastopoulos and B.-L. Hu, Comment on "A Spin Entanglement Witness for Quantum Gravity"; and on "Gravitationally Induced Entanglement between Two Massive Particles is Sufficient Evidence of Quantum Effects in Gravity", 1804.11315.

2. S. Bose, A. Mazumdar, G. W. Morley, H. Ulbricht, M. Toroš, M. Paternostro, A. A. Geraci, P. F. Barker, M. S. Kim and G. Milburn, Spin Entanglement Witness for Quantum Gravity, Physical Review Letters 119 (2017), no. 24 240401 [1707.06050].

3. M. Christodoulou, C. Rovelli. On the possibility of laboratory evidence for quantum superposition of geometries. *Phys.Lett.B*, 2019, 792, pp.64-68.

4. M. J. Hall and M. Reginatto, On two recent proposals for witnessing nonclassical gravity, Journal of Physics A: Mathematical and Theoretical 51 (2018), no. 8 085303 [1707.07974].

5. C. Marletto and V. Vedral, Gravitationally Induced Entanglement between Two Massive Particles is Sufficient Evidence of Quantum Effects in Gravity, Physical Review Letters 119 (2017).

6. C. Marletto and V. Vedral, "Why we need to quantise everything, including gravity", *npj Quantum Information*, **3**, 29, (2016).